\documentclass[12pt]{article}
\usepackage{amsmath}
\usepackage{amscd}
\usepackage{amssymb}
\usepackage{array}

\begin{document}
\rightline{CERN-PH-TH/2004-191}

\rightline{IFIC/04-51}

\rightline{FTUV-04/0929}

\newcommand{\R}{\mathbb{R}}
\newcommand{\C}{\mathbb{C}}
\newcommand{\Q}{\mathbb{Q}}
\newcommand{\Z}{\mathbb{Z}}
\newcommand{\Hb}{\mathbb{H}}

\newcommand{\Ss}{Scherk--Schwarz }
\newcommand{\KK}{ Kaluza--Klein }
\newcommand{\mm}{\mathcal{M}}
\newcommand{\cM}{\mathcal{M}}
\newcommand{\cV}{\mathcal{V}}
\newcommand{\cD}{\mathcal{D}}
\newcommand{\cC}{\mathcal{C}}
\newcommand{\cS}{\mathcal{S}}
\newcommand{\cU}{\mathcal{U}}

\newcommand{\rE}{\mathrm{E}}
\newcommand{\ii}{\mathrm{i}}
\newcommand{\rSp}{\mathrm{Sp}}
\newcommand{\rSO}{\mathrm{SO}}
\newcommand{\rSL}{\mathrm{SL}}
\newcommand{\rSU}{\mathrm{SU}}
\newcommand{\rUSp}{\mathrm{USp}}
\newcommand{\rSpin}{\mathrm{Spin}}
\newcommand{\rU}{\mathrm{U}}
\newcommand{\rF}{\mathrm{F}}
\newcommand{\rGL}{\mathrm{GL}}
\newcommand{\rG}{\mathrm{G}}
\newcommand{\rK}{\mathrm{K}}

\newcommand{\fgl}{\mathfrak{gl}}
\newcommand{\fu}{\mathfrak{u}}
\newcommand{\fsl}{\mathfrak{sl}}
\newcommand{\fsp}{\mathfrak{sp}}
\newcommand{\fusp}{\mathfrak{usp}}
\newcommand{\fsu}{\mathfrak{su}}
\newcommand{\fp}{\mathfrak{p}}
\newcommand{\fso}{\mathfrak{so}}
\newcommand{\fl}{\mathfrak{l}}
\newcommand{\fg}{\mathfrak{g}}
\newcommand{\fr}{\mathfrak{r}}
\newcommand{\fe}{\mathfrak{e}}
\newcommand{\ft}{\mathfrak{t}}
\newcommand{\id}{\relax{\rm 1\kern-.35em 1}}

\vskip 1cm

  \centerline{\LARGE \bf Generalized dimensional  reduction }

 \bigskip

    \centerline{\LARGE \bf  of supergravity with eight supercharges}



\vskip 1.5cm

\centerline{L. Andrianopoli$^{\flat}$,  S.
Ferrara$^{\flat,\sharp}$ and M. A. Lled\'o$^{\natural}$.}
 \vskip
1.5cm

\centerline{\it $^\flat$ Department of Physics, Theory Division}
\centerline{\it
 CERN, CH 1211 Geneva 23, Switzerland.} \centerline{{\footnotesize e-mail:
Laura.Andrianopoli@cern.ch, \; Sergio.Ferrara@cern.ch}}

\medskip

\centerline{\it \it $^\sharp$INFN, Laboratori Nazionali di
Frascati, Italy.}

\medskip
\centerline{\it $^\natural$
 Departament de F\'{\i}sica Te\`orica,
Universitat de Val\`encia and IFIC}
 \centerline{\small\it C/Dr.
Moliner, 50, E-46100 Burjassot (Val\`encia), Spain.}
 \centerline{{\footnotesize e-mail: Maria.Lledo@ific.uv.es}}


\vskip 1.5cm

\begin{abstract}
We describe some recent investigation about the structure of generic $D=4,5$ theories obtained by generalized dimensional reduction
of $D=5,6$ theories with eight supercharges.
We relate the \Ss reduction to a special class of $N=2$ no-scale gauged supergravities.
\end{abstract}

\vskip 1.3cm
------------------------------------------------------------------------------------------
\vskip 5mm

\noindent
{\em Contribution
to the proceedings of ``NathFest'' at PASCOS conference, Northeastern University, Boston, Ma, August 2004
}

 \vfill\eject

\section{Introduction}

We report here on massive deformations of theories with local supersymmetry with eight supercharges, obtained by generalized dimensional reduction
from higher dimensional theories \cite{5to4}.

The case with eight supercharges is of particular interest because such theories have a rich structure of multiplets and interactions,
although the extended supersymmetry considerably limits their geometric structure.

The investigation covers ungauged supergravities in $D=6,5$ compactified with \Ss (SS) phases (duality twists) to $D=5,4$ respectively.

The SS mechanism \cite{ss} relies on the presence of a global symmetry group in the higher dimensional theory.

To obtain a theory with eight supercharges in lower dimensions we could start with a higher dimensional theory with higher supersymmetry and then reduce it by acting,
for example, with some discrete $\Z^p$ symmetry on a torus (orientifold).
Here we limit ourselves to the simpler possibility of starting with a theory with eight supercharges, and this restricts the investigation to $D=6$ and $5$ ungauged
higher dimensional theories, to get theories with massive deformations in $D\geq 4$.  

The global symmetries must act on the fields of the theory which then acquire mass terms upon generalized dimensional reduction
on $S^1$. these symmetries always exist for $D=5,6$ supergravities with $N>2$ local supersymmetry, but they do not generally exist for $N=2$.
We must therefore require that the scalar manifolds of the corresponding multiplets (vector- and hyper-multiplets for $D=5$,  tensor- and hyper-multiplets for $D=6$)
have some isometries.
The only exception to this possibility is the case without hypermultiplets where a mass deformation corresponding to a Fayet-Iliopoulos term, with no contribution to the
 scalar potential, is also possible.

The scalar manifold in a $D=5$ theory with $n_v$ vector multiplets and $n_h$ hypermultiplets is of the type 
$\mathcal{M}_{RS} \times \mm_Q $.
Here $\mm_{RS}$, the manifold spanned by the scalars in the vector multiplets,  is a real manifold with certain restrictions on its geometry, 
described by ``real special'' geometry \cite{gst2}, while $\mm_Q$, the manifold spanned by the scalars in the hypermultiplets, is a quaternionic manifold of dimension $n_h$ (real dimension $4n_h$) \cite{bw}. 

For the $D=6$  $(2,0)$ chiral theories with eight supercharges the scalar manifold is 
$ \frac{\rSO(1,n_T)}{\rSO(n_T)} \times \mm_Q$
where $n_T$ denotes the number of tensor multiplets \cite{ro}  and, as before, $\mm_Q$ is an $n_h$-dimensional quaternionic manifold.
The number $n_V$ of vector multiplets is linked to $n_T , n_h$ by an anomaly cancellation requirement
$$n_h-n_V + 29 n_T =273.$$
In this case the SS phase may act on both types of manifolds, as well as on different multiplets of the theory, depending on the nature of the global symmetry.

Following Scherk and Schwarz, we are going to consider in the following only {\em flat groups} obtained by SS reduction of a $D=5,6$ theory, thus getting, in $D=4,5$, 
a semi-positive definite scalar potential which generally has some flat directions, irrespectively if supersymmetry is broken or not.
Therefore the SS reduction always defines no-scale models \cite{ln}, which were first considered, in the context of $N=1$ supergravity, 
in the early 80's.
No-scale models also appear in warped superstring compactifications where ``internal'' fluxes are turned on. For a review, see \cite{frey}.

At the supergravity level the difference between \Ss and flux compactifications mainly resides in the fact that, when supersymmetry is broken,
massive multiplets are BPS saturated in the former case, while they are long in the latter.

The BPS $\rU(1)$ central charge is gauged by the graviphoton which, together with the other matter gauge fields, defines a {\em non-abelian} gauge
algebra corresponding to a {\em flat group}.

We find \cite{5to4} that the structure of the flat group gauged in $D=4$ ($D=5$)  is universal, in the sense that its structure is common to all $N=2$ models.
It is of the form $\rU(1)\circledS ~  \R^{n_v +1}$, where $n_v$ is the total number of $D=5$ ($D=6$) vector multiplets, and in the general case 
the $\rU(1)$ symmetry, which is gauged by the four (five) dimensional graviphoton $B_\mu$, 
\footnote{By graviphoton we call the $(D-1)$-dimensional vector coming from the $D$-dimensional metric.
This is not the same as the vector partner of the metric in the $N=2$, $D=4,5$ gravity supermultiplet.} acts on both the 
real manifold as well as the quaternionic manifold.

If the $\rU(1)$  has a component on the $\rSU(2)$ R-symmetry of the $D=5$ ($D=6$) theory, then supersymmetry is broken \cite{svn}.
When the quaternionic manifold is involved in the gauging, this is realized when the pull-back on space-time of the $\rSU(2)$ connection $\omega_A^{\ B}$ is non vanishing,
 with in particular
$${\omega_D}_A^{\ B} \neq 0 \, ; \qquad (A,B=1,2); \quad D=5,6 .$$
In absence of hypermultiplets, for supersymmetry to be broken the $\rU(1)$ must have a component on the global $\rSU(2)$ R-symmetry of the $D=5,6$ theory, and this 
originates a $N=2$ Fayet--Iliopoulos term in lower dimensions \cite{dkz}. 

On the other hand,  if no SS phase is introduced in the real-special manifold, then the $\rU(1)$ has no component on the vector multiplet (tensor multiplets)
 directions and
 the flat group is abelian, with all the vectors remaining massless.

 In the general case, the $\rU(1)$ charge has components on the isometries of  both  special and quaternionic manifolds.

We denote by $\Gamma^{ab}= -\Gamma^{ba}$ ($a,b =1,\cdots n_v$) and $\Delta^{\alpha\beta}=\Delta^{\beta\alpha} $  ($\alpha,\beta =1,\cdots 2 n_h$)
 respectively  the $\rSO(n_v)$ ($\rSO(n_T)$) spin-connection one form on the real-special manifold 
and  the $\rUSp(2n_h)$ symplectic connection one-form  on the quaternionic manifold.
 For a Higgs mechanism to take place the following
conditions for the pull-back on 5D (6D) space-time of these connections must be met
$$ {\Gamma_D}^{a b}\neq 0 \,, \quad {\Delta_D}^{\alpha\beta}\neq 0\, . $$
 
Note that, if ${\omega_D}_A^{\ B}=0 $, then supersymmetry remains unbroken and a pure supersymmetric Higgs mechanism occurs.  
Then massive BPS multiplets are generated for both vector- and hypermultiplets in four dimensions.

In section 2 we discuss the SS reduction of $D=5$ and $D=6$ theories with eight supercharges. 
In Section 3 we then give the gauged supergravity interpretation of the reduction.


\section{On the \Ss reduction of the $N=2$ lagrangian from $D$ to $D-1$}
\subsection{General facts}
The generalized \Ss compactification on $S^1$ corresponds  to impose on the fields boundary conditions  (duality twists)
of the type
\begin{equation}
\phi(x,y+2\pi R) = U(y) \phi(x,y) ,
\end{equation}
where $y$ denotes the $D$-th direction on the circle of radius $R$, and $U$ is a rigid symmetry of the $D$ dimensional theory.

For continuous symmetries $U=exp[My]$ where the SS phase $M$ is taken in the compact Cartan subalgebra of the global symmetry group of the $D$ dimensional symmetry.

It is easy to show that the effect of the SS phase is to generate a scalar potential and bilinear fermionic generalized mass terms in the reduced theory, coming merely from terms where a space-time derivative occurs.

Indeed, from a non-linear $\sigma$-model interaction
\begin{equation}
g_{ij}(\phi ) \partial_\mu \phi^i \partial_\nu \phi^j g^{\mu\nu} \sqrt{-g}
\end{equation}
in $D$ dimensions, we retrieve in $D-1$ the scalar potential 
\begin{equation}
V_{D-1}(\phi)= e^{-\frac{2(D-2)}{D-3}\sigma} g_{ij}(\phi ) M^i_{\ l} \phi^l M^j_{\ k} \phi^k  \sqrt{-g_{D-1}}
\end{equation}
where $e^\sigma = \sqrt{g_{DD}}$ and $M^i_{\ l} = - M_l^{\ i}$ is a compact Cartan generator of the global symmetry.
The extrema of the potential are found for 
\begin{equation}
\partial_D \phi^i = M^i_{\ l}\phi^l =0 \label{extr}
\end{equation}
From (\ref{extr}) it then follows that the scalars which are not fixed correspond to the vanishing eigenvalues of the matrix $M$. This
(together with the $\sigma$ direction)  is the residual moduli space of SS compactifications at the classical level.


\subsection{SS reduction of the $D=5,6$ $N=2$ lagrangians}

The basic quantities of the $D=5,6$ lagrangians which become relevant in the discussion of the SS reduction are the kinetic terms for the scalars 
and for the spin $\frac 12 , \frac 32 $ fields, which are related to the scalar potential and the fermionic mass terms of the $D=4,5$ dimensionally reduced theory.

These kinetic terms contain, in the fermionic covariant derivatives, the real geometry spin-connection \cite{dvvp}, 
as well as the symplectic $\rUSp(2n_h)$ and $\rSU(2)$ connections of the quaternionic manifold \cite{bw}.

Upon SS reduction, the pull-back on space-time of such one-form connections contribute the terms
$$ {\Delta_D}_{\alpha\beta}\, ;  \quad {\omega_D}_{AB}\, ; \quad {\Gamma_D}^{ab} \qquad\qquad\qquad ({\omega_D}_{AB}={\omega_D}_{A}^{\ C} \epsilon_{BC}) $$ 
which will determine the quadratic mass terms of the fermions.

Similarly, the $D=4,5$ scalar potential is \cite{svn}
\begin{equation}
V_{D-1}(\sigma , \varphi , q)=  e^{-2\frac{D-2}{D-3}\sigma}\left[\frac 12 P_D^a(\varphi) P_{D a}(\varphi) + 
{\mathcal {U}}_D^{\alpha A}(q)\,{\mathcal {U}}_D^{\beta B}(q)\C_{\alpha\beta}
 \epsilon_{AB}\right]
\label{Dpot}
\end{equation}
where $P_D^a(\varphi) = P^a_{i} \partial_D \varphi^{i}$ and ${\mathcal {U}}_D^{\alpha A}(q)= {\mathcal {U}}_u^{\alpha A} \partial_D q^u$ denote the $D$-th components
 of the pull-back on space-time of the scalar vielbeins of the theory in $D=5,6$. 
${\mathcal {U}}_u^{\alpha A}$ is the vielbein of the quaternionic manifold (with $i=a=1,\cdots n_v$; $A=1,2$; $\alpha = 1,\cdots 2n_h$ and $u=1,\cdots 4n_h$).
 $P^a_{i}$ for $D=5$ is the vielbein of the real-special manifold while for $D=6$ is the vielbein of the symmetric space $\frac{\rSO(1,n_T)}{\rSO(n_T)}$.

It is obvious from (\ref{Dpot}) that the potential $V$ is positive definite and it has an extremum at the points for which
\begin{equation}
P_D^a(\varphi) ={\mathcal {U}}_D^{\alpha A}(q)=0 .
\end{equation}
These are the vacua of the theory.

The equation $P^a_D(\varphi) =0$ can be further specified for D=5 as follows.
Let us call $t^I(\varphi)$ ($I=1,\cdots n_v+1$) the  $D=5$ special coordinates, subject to the constraint \cite{gst2} 
\begin{equation}
t^I t^J t^K d_{IJK}=1 .
\label{constr}
\end{equation}
 They form a representation of the
full duality group of the $D=5$ theory.

Then  the potential can be computed from the real-special geometry relations giving
\begin{equation}
 -\frac 32 e^{-3\sigma} \partial_5 t^I \partial_5 t^J t^K d_{IJK}
\label{veryspec}
\end{equation}
where \begin{equation}
\partial_5 t^I(\varphi) = t^I_{,i} \partial_5 \varphi^i = t^I_{,i} M^i_j \varphi^j .
\label{ssphase}
\end{equation}


\section{Interpretation as gauged $N=2$ supergravity in $D-1$ dimensions}

In the $D=4,5$ framework, mass terms and scalar potential arise from the gauging procedure.
This was fully exploited for the $N=8$ case in \cite{adfl1,adfl2}, and  for the $N=2$ case in \cite{5to4}.
Extended supergravities were further considered from this point of view in \cite{dh,dst,vz}.

In the 4D case, the vectors are $B_\mu$ (the \KK graviphoton) and $Z^I_\mu = A^I_\mu -A^I_5 B_\mu$.
This in particular shows that the five-dimensional graviphoton belongs, together with $\phi_5 =\sqrt{g_{55}}$, to an additional vector multiplet in four dimensions, 
while the $D=4$ graviphoton comes from the \KK vector, corresponding to the decomposition of the five-dimensional  space-time vielbein as
\begin{equation}
\left( \hat V^a_\mu = e^{-\frac\sigma 2} V^a_\mu ; \quad  \hat V^5_\mu = e^{ \sigma }  B_\mu  ; \quad  \hat V^5_5 = e^{ \sigma} \right)
.\end{equation}
This is merely due to the choice of $D=4$ special coordinates $X^\Lambda$ which set $X^0$ to correspond to $B_\mu$. This is not 
the same as the ``free'' supermultiplet assignment.

The very-special manifold $\mathcal{M}_{\C}$ in $D=4$ has $\R^{n_v+1}$ isometries (corresponding to the $A^I_5$ shift invariance of the lagrangian)
\begin{equation}
\delta A_5^I = r^I \, , \qquad I=1,\cdots n_v+1.
\end{equation}
They act on the $n_v+2$ vectors in $D=4$ as follows
\begin{equation}
\delta Z^I_\mu = - r^I B_\mu \, ; \quad \delta B_\mu =0 .
\end{equation}

Let us now consider $B_\mu $ to gauge a $\rU(1)$ group belonging  to the maximal compact subgroup of the isometry group of the very-special  manifold, 
and in particular let us take it
 in its Cartan subalgebra $\mathcal{H}_{\C}$.

If $t^I$ is a representation of $\mathcal{H}_{\C}$, then we may consider the following flat group
\begin{eqnarray}
&\left[ t^I, t^0 \right] & =
\  M^I_J t^J \nonumber\\
&\left[ t^I,t^J \right] & =
\ 0
\end{eqnarray}
that is, by setting $t^\Lambda =(t^0, t^I)$ ,
\begin{equation}
\left[ t^\Lambda, t^\Sigma \right]
= f^{\Lambda\Sigma}_{\ \  \Delta} t^\Delta\, , \qquad \Lambda =(0,I)
\end{equation}
with $f^{I0}_{\ \ J} = M^I_J$, the others vanishing.
 The  $\rU(1)$ isometry gauged by the $B_\mu$ gauge field may have components 
both on the very-special manifold and on the quaternionic manifold, that is \footnote{By $I^Q_{\rUSp(2n_h)}$ and  $I^Q_{\rSU(2)}$ we mean the SS phase with
flattened indices \cite{svn}.}
$$t^0=I^{SG} + I^Q_{\rUSp(2n_h)} + I^Q_{\rSU(2)}.$$ 
The charges corresponding to gauging some special geometry isometries $I^{SG}$  are given by the SS phase $M^I_J$,
while the ones corresponding to gauging quaternionic isometries $I^Q$ are labeled by a matrix $M^u_{\ v}$ \cite{5to4}.

The gauge transformation of the $z^I$ coordinates is holomorphic and has the form
\begin{equation}
\delta z^I = M^I_J \left(z^J \xi^0 - \xi^J\right)
\label{gaugetrscalars}
\end{equation} 
Of course, as the gauge parameters $\xi^\Lambda$ are real, the non-homogeneous part only affects $\Re z^I$.

From the structure constants of the non abelian gauge algebra we also get the gauge transformation of the vectors
\begin{eqnarray}
\delta Z^I_{\mu} &=& \partial_\mu \xi^I + M^I_J \left(\xi^0 Z^J_\mu - \xi^J B_\mu \right)\nonumber\\
\delta B_{\mu} &=& 
\partial_\mu \xi^0
\label{gaugetrvectors}
\end{eqnarray}
and the expression  for the vector field-strengths
\begin{eqnarray}
F^I_{\mu\nu} &=&  \partial_\mu Z^I_\nu -\partial_\nu Z^I_\mu + M^I_J \left(Z^J_\mu B_\nu - Z^J_\nu B_\mu \right)\nonumber\\
B_{\mu\nu} &=& 
\partial_\mu B_\nu -\partial_\nu B_\mu 
.\end{eqnarray}
Note that, because of the gauged translations, a Chern--Simons-like term is present in the $D=4$ lagrangian \cite{dlv}
\begin{equation}
\frac 13 \epsilon^{\mu\nu\rho\sigma} d_{IJK} M^K_L Z^I_\mu Z^L_\nu \partial_\rho Z^J_\sigma .
\end{equation}
It comes by dimensional reduction of the $D=5$ Chern--Simons term \cite{adfl1}
\begin{equation}
d_{IJK} A^I \wedge F^J \wedge F^K .
\end{equation}
Note that $d_{(IJK} M^K_{\ L)}=0$ and that the non-abelian contributions vanish identically \cite{adfl1}.

In the $D=6$ case, the graviton multiplet contains a self dual tensor field, while
the tensors from the tensor multiplets are anti-self dual. We
denote the set of tensor fields as $B^r$, $r=0,\dots n_T$, with
$B^0$ pertaining to the graviton multiplet.

In presence of vector multiplets, the vectors
($A^x$, $x=1,\dots n_V$) couple to the tensor fields  and their
interaction term is of the form \cite{fms,ns2}
$$C_{rxy}B^r\wedge F^x\wedge F^y, \qquad F^x=dA^x,$$  with
$C_{rxy}$=constant. This term is related by supersymmetry to the
kinetic term of the vectors
\begin{equation}
C_{rxy}b^rF^x\wedge ^*\!\!F^y.
\label{kv}
\end{equation}
The fields $b^r$, $r=0,\dots n_T$ satisfy the constraint
$$\eta_{rs}b^rb^s=1,$$
which defines the manifold ${\rSO(1,n_T)}/{\rSO(n_T)}$. The
terms (\ref{kv}) explicitly break the $\rSO(1,n_T)$ symmetry, unless the
vector fields $A^x$ transform under some $n_V$-dimensional
representation $R_V$ of $\rSO(1,n_T)$ with the property that
$\mathrm{Sym}(R_V\otimes R_V)$ contains the vector representation.
In that case, the constants $C_{rxy}$ can be chosen as
invariant couplings. This happens, for instance, if $R_V$ is a
spinor representation of $\rSO(1,n_T)$. Remarkably, this choice
leads after  dimensional reduction on  $S^1$ to the real special
geometries which are homogeneous (in particular, symmetric) spaces
\cite{al,ce,dv,dvvp}.

The reduced 5D theory has a special geometry defined by a cubic polynomial constraint \cite{fms}
$$\mathcal{V} =d_{IJK} t^It^J t^K=1,\qquad I,J,K=1,\dots n_T+n_V+2$$
where 
\begin{equation}
\mathcal{V} = 3\left( z\eta_{rs} b^r b^s + C_{rxy}~ b^r a^x
a^y\right) ~; \quad  r=0,1,\cdots n_T ~; \quad x=1,\cdots n_V.
\label{cubic}
\end{equation}
$\eta_{rs}$ is the $(1,n_T)$ Lorentzian metric related to the
space $\rSO(1,n_T)/\rSO(n_T)$ (parametrized by $b^r$), $z=\sqrt{g_{66}}=e^\sigma $ is the KK scalar and $a^x=A^x_6$
are the axions.

Under the above assumption, the SS reduction produces a theory with a
flat gauge group of the form $\rU(1)\ltimes R_V$, where the
$\rU(1)$ generator  is in the Cartan subalgebra (CSA) of the
maximal compact subgroup $\rSO(n_T)$ of the global symmetry
$\rSO(1,n_T)$. The $\rU(1)$ group is gauged by the vector coming
from the metric in dimension six. The tensors are in a vector
representation of $\rSO(1,n_T)$, so they are charged under U(1)
(except for some singlets as $B^0$).

We remark that in order to introduce a SS phase in the
tensor-vector multiplet sector it is actually sufficient that the
constants $C_{rxy}$ preserve a $U(1)$ subgroup of $\rSO(1,n_T)$,
which is a much weaker assumption.

The generator of the group U(1) may also have a component on the
isometries of the quaternionic manifold \cite{5to4}; in particular,
it may have a component in the CSA of the SU(2) R-symmetry, then
breaking supersymmetry (notice that this can happen even if
hypermultiplets are not present, corresponding to a $D=5$
Fayet-Iliopoulos term). The SS reduction leads to a positive
semidefinite potential  also in this case. The $D=5$
interpretation of the theory must correspond to a gauging with the
term $V_R=0$ (see Ref.
\cite{gst}).

In Ref. \cite{5to4} this analysis has been applied to all $\sigma$-models based on symmetric real and quaternionic geometries.
For this case a detailed study of the SS mass spectrum and of the residual moduli space is explicitely performed.


\section*{Acknowledgements}

 The work of S.F. has been supported in
part by the D.O.E. grant DE-FG03-91ER40662, Task C, and in part by
the European Community's Human Potential Program under contract
HPRN-CT-2000-00131 Quantum Space-Time, in association with INFN
Frascati National Laboratories.

The work of M. A. Ll. has been supported by the research grant BFM
2002-03681 from the Ministerio de Ciencia y Tecnolog\'{\i}a
(Spain) and from EU FEDER funds.

\end{document}